\documentclass[twocolumn,twoside,slac]{revtex4}
\usepackage{graphicx}
\usepackage{fancyhdr}
\begin{document}

  \title{R-GMA: First results after deployment}

  \author{Andrew Cooke} 
  \author{Werner Nutt}
  \affiliation{Heriot-Watt, Edinburgh, UK}

  \author{James Magowan} 
  \author{Manfred Oevers}
  \author{Paul Taylor}
  \affiliation{IBM-UK}

  \author{Ari Datta}
  \author{Roney Cordenonsi}
  \affiliation{Queen Mary, University of London, UK}

  \author{Rob Byrom} 
  \author{Laurence Field}
  \author{Steve Hicks}
  \author{Manish Soni}
  \author{Antony Wilson}
  \author{Xiaomei Zhu}
  \affiliation{ PPARC, UK}

  \author{Linda Cornwall}
  \author{Abdeslem Djaoui}
  \author{Steve Fisher}
  \affiliation{Rutherford Appleton Laboratory, UK}

  \author{Norbert Podhorszki}
  \affiliation{SZTAKI, Hungary}

  \author{Brian Coghlan}
  \author{Stuart Kenny}
  \author{David O'Callaghan}
  \author{John Ryan}
  \affiliation{Trinity College Dublin, Ireland}

  \begin{abstract}
    We describe R-GMA (Relational Grid Monitoring Architecture) which
    is being developed within the European DataGrid Project as an Grid
    Information and Monitoring System. Is is based on the GMA from
    GGF, which is a simple Consumer-Producer model. The special
    strength of this implementation comes from the power of the
    relational model. We offer a global view of the information as if
    each VO had one large relational database. We provide a number of
    different Producer types with different characteristics; for
    example some support streaming of information. We also provide
    combined Consumer/Producers, which are able to combine information
    and republish it. At the heart of the system is the mediator,
    which for any query is able to find and connect to the best
    Producers to do the job. We are able to invoke MDS info-provider
    scripts and publish the resulting information via R-GMA in
    addition to having some of our own sensors. APIs are available
    which allow the user to deploy monitoring and information services
    for any application that may be needed in the future. We have used
    it both for information about the grid (primarily to find what
    services are available at any one time) and for application
    monitoring. R-GMA has been deployed in Grid testbeds, we describe
    the results and experiences of this deployment.
  \end{abstract}

  \maketitle

  \thispagestyle{fancy}

  \section{Introduction}

\begin{figure}[htfb]
    \includegraphics[scale=0.4]{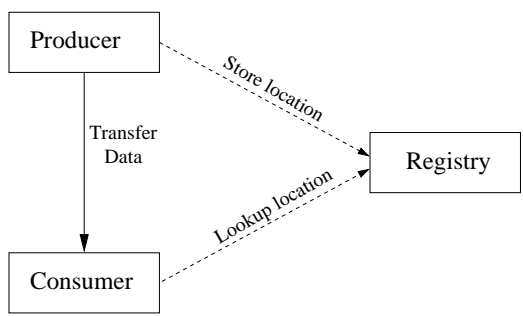}
    \caption{Grid Monitoring Architecture}
    \label{figure-gma}
  \end{figure}

  The Grid Monitoring Architecture (GMA)\cite{ref:perf-arch} of the
  GGF, as shown in Figure~\ref{figure-gma}, consists of three
  components: \textit{Consumers}, \textit{Producers} and a directory
  service, which we prefer to call a \textit{Registry}).

  In the GMA Producers register themselves with the Registry and
  describe the type and structure of information they want to make
  available to the Grid. Consumers can query the Registry to find out
  what type of information is available and locate Producers that
  provide such information. Once this information is known the
  Consumer can contact the Producer directly to obtain the relevant
  data.  By specifying the Consumer/Producer protocol and the
  interfaces to the Registry one can build inter-operable
  services. The Registry communication is shown on
  Figure~\ref{figure-gma} by a dotted line and the main flow of data
  by a solid line.

  The current GMA definition also describes the registration of
  Consumers, so that a Producer can find a Consumer. The main reason
  to register the existence of Consumers is so that the Registry can
  notify them about changes in the set of Producers that interests
  them.

  The GMA architecture was of course devised for monitoring but we
  think it makes an excellent basis for a \emph{combined} information and
  monitoring system. We have argued
  before\cite{ref:time-bncod-2001} that the only thing which
  characterises monitoring information is a time stamp, so we insist
  upon a time stamp on all measurements - saying that this is the time
  when the measurement was made, or equivalently the time when the
  statement represented by the tuple was true.

  The GMA does not constrain any of the protocols nor the underlying
  data model, so we were free when producing our implementation to
  adopt a data model which would allow the formulation of powerful
  queries over the data.

  R-GMA is a relational implementation of the GMA which brings the
  power and flexibility of that model. R-GMA creates the impression
  that you have one RDBMS per VO. However it is important to
  appreciate that what our system provides is a way of using the
  relational model in a Grid environment and that we have \emph{not}
  produced a general distributed RDBMS. All the producers of
  information are quite independent. It is relational in the sense
  that Producers announce what they have to publish via an SQL CREATE
  TABLE statement and publish with an SQL INSERT and that Consumers
  use an SQL SELECT to collect the information they need.

  R-GMA is built using servlet technology and is being migrated
  rapidly to web services and specifically to fit into an
  OGSA\cite{ref:ogsa-spec} framework.

  \section{Query types and Producer Types}
  
  We have so far defined not just a single Producer but five different
  types: a DataBaseProducer, a StreamProducer, a ResilientProducer, a
  LatestProducer and a CanonicalProducer. All appear to be Producers
  as seen by a Consumer - but they have different characteristics.
  The CanonicalProducer, though in some respects the most general, is
  somewhat different as there is no user interface to publish data via
  an SQL \texttt{INSERT} statement. Instead it triggers user code to
  answer an SQL query. The other Producers are all
  \texttt{Insertable}; this means that they all have an interface
  accepting an SQL \texttt{INSERT} statement.

  The other producers are instantiated and given the description of
  the information they have to offer by an SQL \texttt{CREATE TABLE}
  statement and a \texttt{WHERE} clause expressing a predicate that is
  true for the table. Currently this is of the form \texttt{WHERE
  (column\_1=value\_1 AND column\_2=value\_2 AND ...)}.  To publish
  data, a method is invoked which takes the form of a normal SQL
  \texttt{INSERT} statement.

  Three kinds of query are supported: History, Latest and
  Continuous. The history query might be seen as the more traditional
  one, where you want to make a query over some time period -
  including.``all time''. The latest query is used to find the current
  value of something and a continuous query provides the client with
  all results matching the query as they are published. A continuous
  query is therefore acting as a filter on published data.

  The DataBaseProducer supports history queries. It writes each record
  to an RDBMS. This is slow (compared to a StreamProducer) but it can
  handle joins.  The StreamProducer supports continuous queries and
  writes information to a memory structure where it can be picked up
  by a Consumer. The ResilientStreamProducer is similar to the
  StreamProducer but information is backed up to disk so that no
  information is lost in the event of a system crash. The
  LatestProducer supports latest queries by holding only the latest
  records in an RDBMS.

  Latest records are defined in terms of something similar to a
  primary key. Each record has a time stamp, one or more fields which
  define what is being measured (e.g. a hostname) and one or more
  fields which are the measurement (e.g. the 1 minute CPU load
  average). The time stamp and the defining fields are close to being
  a primary key - but as there is no way of knowing who is publishing
  what across the Grid, the concept of primary key (as something
  globally unique) makes no sense. The LatestProducer will replace an
  earlier record having the same defining fields, as long as the time
  stamp on the new record is more recent, or the same as the old one.

  Producers, especially those using an RDBMS, may need cleaning from
  time to time.  We provide a mechanism to specify those records of a
  table to delete by means of a user specified SQL \texttt{WHERE}
  clause which is executed at intervals which are also specified by
  the user.  For example it might delete records more than a week old
  from some table or it may only hold the newest one hundred rows, or
  it might just keep one record from each day.

\begin{figure}[htfb]
    \includegraphics[scale=0.5]{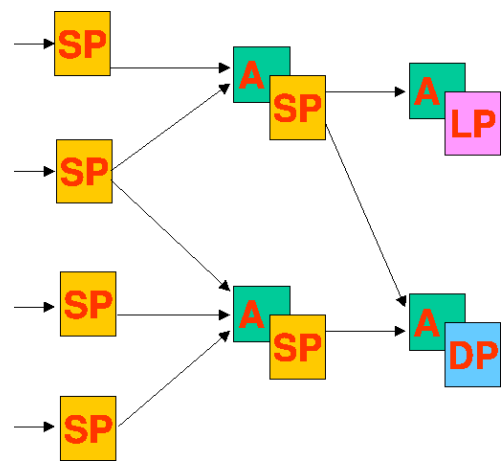}
    \caption{A possible topology of R-GMA components}
    \label{figure-topo}
  \end{figure}

  Another valuable component is the Archiver which is a combined
  Consumer-Producer. You just have to tell it what to collect and it
  does so on your behalf.  An Archiver works by taking over control of
  an existing Producer and instantiating a Consumer for each table it
  is asked to archive. This Consumer then connects via the mediator to
  all suitable Producers and data starts streaming from those
  Producers, through the Archiver and into the new Producer. The
  inputs to an Archiver are always streams from a StreamProducer or a
  ResilientStreamProducer.  It will re-publish to any kind of
  ``Insertable''. This allows useful topologies of components to be
  constructed such as the one shown in Figure~\ref{figure-topo}.

  This shows a number of StreamProducers (labelled SP) which is
  normally the entry point to R-GMA. There is then a layer of
  Archivers (A) publishing to another StreamProducer. Finally there is
  an Archiver to a LatestProducer (LP) and an Archiver to a
  DataBaseProducer (DP) to answer both Latest and History queries.

  We intend to allow some kinds of producer to answer more than one
  kind of query - but for now we are keeping it simple.

  \section{Applications of R-GMA}
  R-GMA has applications right across the Grid. First it can be used
  as a replacement for MDS. A small tool (GIN) has been written to
  invoke the MDS-like EDG info-providers and publish the information
  via R-GMA. The info-provider is a small script which can be invoked
  to produce information in LDIF format. All our information providers
  conform to the GLUE (http://www.cnaf.infn.it/~sergio/datatag/glue/)
  schema. Another tool (GOUT) is available to republish R-GMA data to
  an LDAP server for the benefit of legacy applications. However we
  expect that most applications will wish to benefit from the power of
  relational queries. GOUT is an Archiver with a Consumer which
  periodically publishes to an LDAP database. The GIN-GOUT combination
  is not efficient - but it works. Both GIN and GOUT are driven by
  configuration files which define the mapping between the LDAP schema
  and the relational schema.

  R-GMA is also being used for network monitoring where the
  flexibility of the relational model offers a more natural
  description of the problem.

  It is also being used to locate replica catalogs, and to publish
  information to two tables, the Service table and the ServiceStatus
  table. A service publishes its existence when it starts up with an
  entry in the Service table. It does this using a StreamProducer. An
  Archiver to a LatestProducer is instantiated to collect all Service
  information in one place. There are also processes which check the
  functioning of a service and publish the status frequently to the
  ServiceStatus table. This is also published via a StreamProducer and
  is collected by the same Archiver that is used for archiving the
  Service table. So the Service table says what should exist and the
  ServiceStatus gives the current state Grid wide.

  GRM was written for monitoring parallel
  applications\cite{ref:sztaki-cluster-monitoring-2000} where it
  writes logging information to a local file. This has recently been
  modified to make use of R-GMA for transport.

  In addition CMS have adapted their BOSS system which previously
  wrote to a well known RDBMS to simply publish the job status
  information via R-GMA. This BOSS work is reported at this
  conference.

  \section{Tools}
  There are a number of tools available to query R-GMA Producers.
  There is a command line tool, a Java application: Pulse, and the
  R-GMA Browser, which is accessible from a Web browser without any
  R-GMA installation. The Browser offers a few custom queries, and
  makes it easy for you to write your own.

  The command line tool, which is written in Python, is the most
  powerful.  It is designed to do simple things very easily - but if
  you want to do more complex things you must code them yourself. It
  supports one instance of each kind of producer and one archiver at
  any one time. You can also find what tables exist, find details of a
  table and issue any kind of query.

  \section{The registry and the mediator}
  The registry stores information about all producers currently
  available. Currently there is only one physical Registry per
  VO. This bottleneck and single point of failure is being
  eliminated. Code is being written to allow multiple copies of the
  registry to be maintained. Each one acts as master of the
  information which was originally stored in that Registry instance
  and has copies of the information from other Registry
  instances. Synchronisation is carried out frequently. Currently VOs
  are disjoint, we plan to allow information to be published to a set
  of VOs.

  The mediator (which is hidden behind the Consumer interface) is the
  component which makes R-GMA easy to use.  Producers are associated with
  views on a virtual data base. Currently views have the form:

  \begin{quote}
    SELECT * FROM $<$table$>$ WHERE $<$predicate$>$
  \end{quote}

  This view definition is stored in the Registry.  When queries are
  posed, the Mediator uses the Registry to find the right Producers and
  then combines information from them.

  \section{Architecture}
  \begin{figure*}[htfb]
      \includegraphics[scale=0.46]{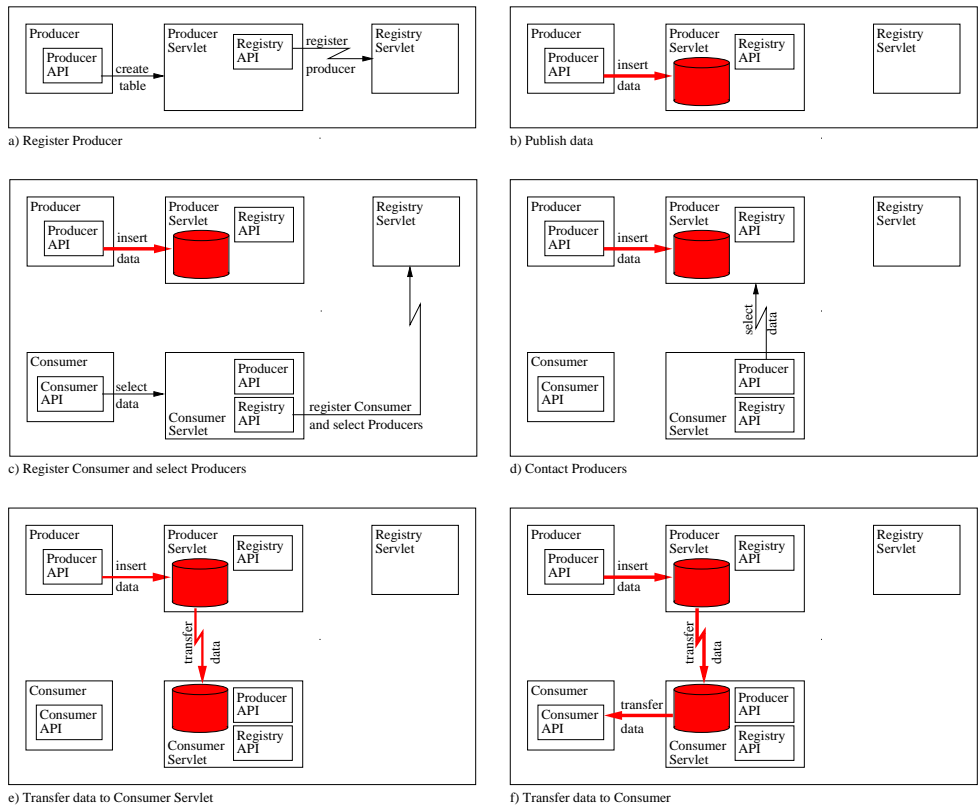}
      \caption{Relational Grid Monitoring Architecture}
    \label{figure-rgma}
  \end{figure*}

  R-GMA is currently based on Servlet technology. Each component has
  the bulk of its implementation in a Servlet.  Multiple APIs in Java,
  C++, C, Python and Perl are available to communicate with the
  servlets. The basic ones are the Java and C++ APIs which are
  completely written by hand. The C API calls the C++ and the Python
  and Perl are generated by SWIG. We make use of the Tomcat Servlet
  container. Most of the code is written in Java and is therefore
  highly portable. The only dependency on other EDG software
  components is in the security area.

  Figure~\ref{figure-rgma} shows the communication between the APIs
  and the Servlets.  When a Producer is created its registration
  details are sent via the Producer Servlet to the Registry
  (Figure~\ref{figure-rgma}a). The Registry records details about the
  Producer, which include the description and view of the data
  published, \emph{but not the data itself}.  The description of the
  data is actually stored as a reference to a table in the Schema. In
  practise the Schema is co-located with the Registry.
  Then when the Producer publishes data, the data are transferred to a
  local Producer Servlet (Figure~\ref{figure-rgma}b).

  When a Consumer is created its registration details are also sent to
  the Registry although this time via a Consumer Servlet
  (Figure~\ref{figure-rgma}c). The Registry records details about the
  type of data that the Consumer is interested in.  The Registry then
  returns a list of Producers back to the Consumer Servlet that match
  the Consumers selection criteria.

  The Consumer Servlet then contacts the relevant Producer Servlets to
  initiate transfer of data from the Producer Servlets to the Consumer
  Servlet as shown in \mbox{Figures~\ref{figure-rgma}d-e}.

  The data are then available to the Consumer on the Consumer Servlet,
  which should be close in terms of the network to the Consumer
  (Figure~\ref{figure-rgma}f).

  As details of the Consumers and their selection criteria are stored
  in the Registry, the Consumer Servlets are automatically notified
  when new Producers are registered that meet their selection
  criteria.

  The system makes use of soft state registration to make it
  robust. Producers and Consumers both commit to communicate with
  their servlet within a certain time. A time stamp is stored in the
  Registry, and if nothing is heard by that time, the Producer or
  Consumer is unregistered. The Producer and Consumer servlets keep
  track of the last time they heard from their client, and ensure that
  the Registry time stamp is updated in good time.

  \section{Results so far}
  Unfortunately we have few results to offer at this stage. It has
  taken some time to get from the state of having something which
  passes all its unit tests (about 400 for the Java API) to a stable
  distributed system - which we think we now have. We have just
  started running performance tests to understand the behaviour of the
  code.  We have so far tested with many Producers, and one Archiver
  feeding into a LatestProducer which is then queried to make sure
  that the Archiver is keeping up with the total flow of data. The
  Producers are publishing data following the pattern expected of a
  ``typical'' site having an SE (Storage Element) and 3 CEs (Computing
  Elements). We found that we were able to support around 40 to 50
  sites publishing data every 30 seconds. 

  A producer is able to publish individual tuples or a vector of
  tuples. Changing this buffering had little effect on the maximum
  number of sites the system is able to support.  We will use tools to
  analyse the code to understand precisely where the bottlenecks are
  occurring - in particular to find out why buffering did not have the
  beneficial effect we expected.

  In general we expect to be able to do better by modifying tomcat
  settings, the virtual machine settings and getting more physical
  memory.  To achieve better performance we may need a layer of
  Archivers combining streams into bigger streams so as to limit the
  fan-in to any one node. The other way to obtain significantly better
  performance is not to attempt to get all the information into one
  place. As the mediator becomes more powerful, it will be able to make
  use of multiple LatestProducer archives, and carry out a distributed
  query over them.

  The effort involved in making meaningful measurements on such a
  system as R-GMA should not be underestimated!
 
  \section{Conclusion}
  We have a useful architecture and an effective implementation with a
  number of components which work well together. We hope that R-GMA
  will have a long, happy and useful life, both in its current form
  and when reincarnated within an OGSA framework. For more details of R-GMA,
  please see: http://hepunx.rl.ac.uk/edg/wp3/

  \begin{acknowledgments}
    The authors wish to thank our patient users, the EU and our national funding agencies.
  \end{acknowledgments}

\end{document}